\documentclass[%
superscriptaddress,
preprint,
 amsmath,amssymb,
 aps,
]{revtex4-2}

\usepackage[english]{babel}
\usepackage{graphicx, grffile}
\usepackage{dcolumn}
\usepackage{physics}
\usepackage{bm}
\usepackage{hyperref}
\usepackage{cleveref}
\usepackage{fancyhdr}
\usepackage{romannum}
\usepackage{enumerate}
\usepackage{natbib}
\usepackage{tabularx}
\usepackage{floatrow}
\usepackage{subfig}
\usepackage{caption}
\usepackage{setspace}
\usepackage{comment}

\newcommand{\myref}[1]{(\ref{#1})}

\graphicspath{{figure/}}

\begin{document}

\renewcommand{\thepage}{\arabic{page}}

\title{Generalized dimension reduction approach for heterogeneous networked systems with time-delay}

\author{Cheng Ma}
\affiliation{Department of Physics, Applied Physics and Astronomy, Rensselaer Polytechnic Institute, \\
Troy, NY 12180, USA }
\affiliation{Network Science and Technology Center, Rensselaer Polytechnic Institute, Troy, NY 12180, USA}
\author{Gyorgy Korniss}
\affiliation{Department of Physics, Applied Physics and Astronomy, Rensselaer Polytechnic Institute, \\
Troy, NY 12180, USA }
\affiliation{Network Science and Technology Center, Rensselaer Polytechnic Institute, Troy, NY 12180, USA}
\author{Boleslaw K. Szymanski}
\affiliation{Department of Physics, Applied Physics and Astronomy, Rensselaer Polytechnic Institute, \\
Troy, NY 12180, USA }
\affiliation{Network Science and Technology Center, Rensselaer Polytechnic Institute, Troy, NY 12180, USA}
\affiliation{Department of Computer Science, Rensselaer Polytechnic Institute, Troy, NY 12180, USA}

\author{Jianxi Gao}
\affiliation{Network Science and Technology Center, Rensselaer Polytechnic Institute, Troy, NY 12180, USA}
\affiliation{Department of Computer Science, Rensselaer Polytechnic Institute, Troy, NY 12180, USA}

\begin{abstract}
	Networks of interconnected agents are essential to study complex networked systems' state evolution, stability, resilience, and control. Nevertheless, the high dimensionality and nonlinear dynamics are vital factors preventing us from theoretically analyzing them. Recently, the dimension-reduction approaches reduced the system's size by mapping the original system to a one-dimensional system such that only one effective representative can capture its macroscopic dynamics. However, the approaches dramatically fail as the network becomes heterogeneous and has multiple community structures. Here, we bridge the gap by developing a generalized dimension reduction approach, which enables us to map the original system to a 
 $m$-dimensional system that consists of $m$ interacting components. Notably, by validating it on various dynamical models, this approach accurately predicts the original system state and the tipping point, if any. Furthermore, the numerical results demonstrate that this approach approximates the system evolution and identifies the critical points for complex networks with time delay. 
	\begin{description}
		\item[Keywords] Generalized dimension reduction, networked systems, tipping point, time-delay
	\end{description}
\end{abstract}

\maketitle

\section{\label{sc:introduction}Introduction}

The networked dynamical systems have been broadly used to study a variety of real-world systems, including ecological webs, social networks, neuronal systems, and infrastructure networks, and many of them consist of numerous components connected via complicated interactions \cite{barabasi2016network, m.e.j.newmanNetworks2018, havlinChallengesNetworkScience2012, gao2022introduction}. Advances in the understanding of system synchronization \cite{strogatzKuramotoCrawfordExploring2000}, dynamical spreading \cite{pastor-satorrasEpidemicSpreadingScaleFree2001}, and catastrophic shifts \cite{schefferCatastrophicShiftsEcosystems2001, lentonTippingElementsEarth2008} have offered important clues on the relationship between complex network topology and dynamical evolution \cite{barzelUniversalityNetworkDynamics2013}. To quantitatively investigate those collective phenomena, we often need to obtain the system's long-term evolution and equilibrium states, if any. However, as the number of components and the system complexity increase, such tasks become increasingly computationally expensive.

One common strategy to address this issue is to use the dimension-reduction approach to reduce the number of components in the system, allowing the construction of a simplified version approximating the original system. Note that in this paper, we use the term ``dimension" to refer to the system size (as commonly used in control theory), i.e., the number of nodes in a networked system. The challenge is to find ways to preserve key properties of the original system and make it computationally manageable. 
In recent years, several dimension-reduction theories and techniques have emerged. 
Pecora {\it et.al.} \cite{pecoraClusterSynchronizationIsolated2014} and Schaub {\it et.al.} \cite{schaubGraphPartitionsCluster2016} inspected network symmetry and proposed the irreducible representation to group nodes into different clusters based on their topology, which provides an exact solution to the original system, enabling the study of cluster synchronization and desynchronization. The limitation is that this framework is only applicable to networks that are extremely sparse or dense, which have nontrivial subgroups by symmetry clustering. 
Gao {\it et.al.} \cite{gaoUniversalResiliencePatterns2016,liu2022network} proposed a degree-weighted mean-field approach that effectively maps any $N$-dimensional network to its one-dimensional one. This approach was used to predict the universal resilience pattern under various types of perturbations. Later, Laurence {\it et.al.} \cite{laurenceSpectralDimensionReduction2019} introduced an eigenvector-based reduction framework from the perspective of spectral analysis, opening up new perspectives on the relationship between the network topology and the approximation accuracy.
Pereira {\it et.al.} \cite{erogluRevealingDynamicsCommunities2020, pereiraHeterogeneouslyCoupledMaps2020} applied mean-field theory to build effective networks from observations and was able to identify the interaction structure, reproduce the behavior of the original network, and predict the critical transitions.
Tu {\it et.al.} \cite{tu2021dimensionality} developed an analytical framework for systems with node-specific dynamics, which collapses complex $N$-dimensional networked system into an $S+1$-dimensional manifold as a function of $S$ effective control parameters, where $ S \ll N $. 
Jiang {\it et.al.} \cite{jiangPredictingTippingPoints2018} and Zhang {\it et.al.} \cite{zhang2022estimating} proposed dimensional reduction approaches to predict the tipping points in bipartite mutualistic dynamics. 
Naseri {\it et.al.} \cite{naseriConvertingHighdimensionalComplex2022} focused on preserving synchronization in dimension-reduced systems by using the eigenvalue decomposition method and Gram-Schmidt orthogonalization.

Those dimension-reduction approaches simplify the analysis of system states for large-scale networks, but most of them rely on the assumption that node degrees are homogeneous (i.e., with low-degree variance), as has been shown in \cite{gaoUniversalResiliencePatterns2016, laurenceSpectralDimensionReduction2019}.
Unfortunately, such a requirement does not hold in many real-world systems, especially those whose underlying networks have scale-free (SF) degree distributions or networks with multiple communities. For these networks, the one-dimensional reduction approach cannot capture the wide range of node activities, especially when the system approaches the critical thresholds of phase transition. The ubiquity of SF networks in the real world \cite{barabasiEmergenceScalingRandom1999, barabasiScaleFreeNetworks2003} requires the generalization of the dimension-reduction framework to heterogeneous networks and networks with multiple communities.
Recently, Vegu\'e {\it et.al.} \cite{vegueDimensionReductionDynamics2023} developed a dimension-reduction approach for modular networks based on the spectral graph theory. The nodes are first classified into $n$ communities according to modular structures, and the community state is then represented by one observable through spectral analysis of a series of matrices and appropriate approximations.  

Delayed interactions are very common in many dynamical systems in science, and engineering \cite{olfati-saberConsensusCooperationNetworked2007, erneuxAppliedDelayDifferential2009, ottoNonlinearDynamicsDelay2019} ranging from climate modeling \cite{keaneClimateModelsDelay2017} and population dynamics \cite{mayTimeDelayStabilityPopulation1973, kuangDelayDifferentialEquations1993} to transportation systems \cite{bandoDynamicalModelTraffic1995} and supply chains \cite{riddallsStabilitySupplyChains2002}. 
In contrast to single-variable systems  \cite{hutchinsonCircularCausalSystems1948, mayTimeDelayStabilityPopulation1973, ruanAbsoluteStabilityConditional2001, ruanDELAYDIFFERENTIALEQUATIONS2006}, stability analysis, consensus formation, or control in multi-component systems with time delays and possible nonlinearities present significant computational challenges as the system size increases \cite{olfati-saberConsensusCooperationNetworked2007,sipahiStabilityStabilizationSystems2011,Sipahi_PRE2012}.
Since time delays experienced by individuals have profound effects on other participants in complex interconnected systems, the insights gained from low-dimensional systems may not be directly applicable to systems with many components. This has been shown by the effects of times delays on consensus formation \cite{olfati-saberConsensusCooperationNetworked2007, sipahiStabilityStabilizationSystems2011,Sipahi_PRE2012} and noisy network synchronization problems \cite{huntNetworkSynchronizationNoisy2010, huntNetworkCoordinationSynchronization2012, huntImpactCompetingTime2011, huntExtremeFluctuationsStochastic2015}. Therefore, one needs to investigate the complex interaction topology to understand the system's evolution and stability.

We proposed a generalized dimension-reduction approach for heterogeneous complex networks and networks with multiple communities. This approach consists of two steps. First, the original network of $N$ nodes is clustered into $m$ clusters of nodes with similar degrees based on the assumption that node states are highly correlated with node degrees (supported by \cite{gaoUniversalResiliencePatterns2016, sanhedraiRevivingFailedNetwork2022}). The weighted average state of each cluster is obtained by the mean-field theory in \cite{gaoUniversalResiliencePatterns2016}, causing the dimension-reduced system to follow the identical evolution mechanism. As a result, it can be described by only $m$ representatives, where $m \ll N$. The system evolution can be well approximated with just a few representatives for systems with various dynamics and a wide range of degree distributions. We find that the number of required clusters peaks when the system approaches the tipping point. More importantly, for the first time, we demonstrate that our generalized dimension reduction approach theoretically enables us to analyze the system with time delays. Therefore, it reduces computational complexity while preserving the essential information of time-delayed system structures, enabling us to convert systems of any size to a low-dimensional version before we apply the corresponding theoretical frameworks.

\section{\label{sec:mathematical framework}Results}

\noindent
\subsection{\label{subsec:networks}Generalized dimension reduction approach}

Let us consider a networked system consisting of $N$ nodes where the evolution of node activities, $\bm{x} = (x_1, x_2, ..., x_N)^T$, can be generally described by Eq.~\myref{eq:multi-variable}. In this manuscript, the term ``dimension" is used to refer to ``the number of nodes" in the original system or ``the number of clusters" in the simplified system. The time-dependent state of node $i$ is represented by a scalar variable $x_i(t) \in R$. Each node follows the same self-dynamics $F(x_i)$ and is also influenced by its neighboring node $j$ according to the pairwise interaction $G(x_i, x_j)$. The element of the adjacency matrix $A$ is either $0$ or $1$, indicating the connectivity between components and the network topology. The edge weight $w$ represents the uniform interaction strength of the entire system.  
	\begin{equation}
		\dv[]{x_i}{ t}  = F(x_i) + w \sum _{j=1}^{N} A_{ij} G(x_i, x_j) \label{eq:multi-variable},
	\end{equation}
To analyze the system evolution for a given dynamics and network topology, one can numerically solve the coupled ordinary differential equations for all nodes according to Eq.~\myref{eq:multi-variable}. However, this approach is often computationally expensive or even infeasible for exceptionally large-scale networks. It is reasonable to assume that nodes of similar topological features have similar evolution patterns. Therefore, one may use only a small number of representatives to describe the system state. This is the main principle behind the dimension-reduction approach to tame the computational cost. This strategy has been proven effective by the one-dimension reduction framework \cite{gaoUniversalResiliencePatterns2016}. For homogeneous networks where all nodes share similar connectivity properties, one observable can represent the average state of the entire system by the appropriate choice of reduction approaches. However, for networks with heterogeneous connectivity, we cannot expect that one representative is sufficient to describe the entire system state, as the node states are much more diverse than homogeneous networks. In this study, we generalize the one-dimension reduction framework by introducing more than one observable, making it applicable to heterogeneous networks.

Because of the high correlation between node states and node degrees \cite{barzelUniversalityNetworkDynamics2013, sanhedraiRevivingFailedNetwork2022}, we partition the network into multiple clusters based on node degrees such that nodes of similar degrees are in the same cluster and their average state is represented by one observable. Therefore, one only needs to numerically solve the dimension-reduced networks with a few number of components. 
First, we obtain the node features from node degrees distribution. For SF networks where node degrees follow a power-law distribution, we take the logarithm of the node degree $k_i$ and define the normalized variable $v_i = \frac{\log(k_i)} { \log(k_{\mathrm{max}})}$ as the node feature, whereas for homogeneous ER networks, we normalize the node degree $v_i = \frac{k_i} { k_{\mathrm{max}}}$ as the node feature.  Based on the feature $v_i$, the $K$-means clustering algorithm is applied to partition the network into $m$ clusters such that nodes in the same cluster have similar node features, including node degrees.
The next step is to construct $m$ macroscopic variables and the corresponding parameters for each cluster state. The operator $\mathcal{L}^{(a)} (\bm{x}) \quad (a=1, 2, ..., m)$ is introduced in Eq.~\myref{eq:L} to calculate the average state of the cluster $a$, $y^{(a)} = \mathcal{L}^{(a)}(\bm{x})$, and the average interaction strength $\beta_{ab} = \mathcal{L}^{(a)} (\bm{k}^{(b)}) $ from the cluster $g_b$ to the cluster $g_a$. Here, $\bm{k} ^{(b)}$ is a vector of incoming connections from the cluster $g_{b}$.

\begin{equation}
	\mathcal{L}^{(a)} (\bm{x}) = \frac{ \bm{u}^{{(a)}^T}\bm{K} \bm{x}  }{ \bm{u}^{{(a)}^T} \bm{A} \bm{1}}  = \frac{ \sum _{i=1, i\in g_a}^{N}k_i x_i}{ \sum _{i=1, i\in g_a}^{N}k_i}. \label{eq:L}
\end{equation}

In Eq.~\myref{eq:L}, $\bm{K}$ is a diagonal matrix with the element $K_{ii} = k_i$ representing the degree of node $i$, $\bm{1} = (1, 1, ..., 1)^T$ is the unit vector, and $\bm{u}^{(a)} = (u^{(a)}_1, u^{(a)}_2, ..., u^{(a)}_N)^T$ is a binary vector indicating that the node $i$ belongs to the cluster $ g_{a}$ if $u^{(a)}_i=1$, otherwise $u^{(a)}_i=0$.
Thus, the operator $\mathcal{L}^{(a)}$ takes the degree-weighted average of the quantity $x_i$ over all nodes in the cluster $g_a$.
By applying the operator $\mathcal{L}^{(a)}$ to Eq.~\myref{eq:multi-variable} and assuming $\mathcal{L}^{(a)} (f(\bm{x})) \approx f( \mathcal{L}^{(a)}(\bm{x}))$, one can derive the dimension-reduction framework defined by Eq.~\myref{eq:m-variable}, where $y^{(a)}$ represents the average state of the cluster $g_a$. (For further details on analytical derivations and numerical verifications, please refer to Supplementary Note II.) Compared with Eq.~\myref{eq:multi-variable}, the dynamics of the simplified system preserves the form of the original dynamics, and only the interaction topology changes. This framework maps the dynamics of the $N$-dimensional complex system into an effective $m$-dimensional equation, and such mapping can significantly reduce the computational cost and enable the use of theoretical tools developed for low-dimensional systems. 

			\begin{equation}
				\dv[]{y^{(a)}}{t}  = F(y^{(a)})+ w \sum _{b=1}^{m}  \beta_{ab} G(y^{(a)}, y^{(b)}) \label{eq:m-variable}
			\end{equation}

			 We use the cluster state $y^{(a)}$ to approximate the state of each node in this cluster, ($x_i = y^{(a)}, i \in g_a$), and thus all node states $\bm{x}$ are accessible by the dimension-reduced system. To understand the overall evolution at the system level, we define the operator $ \mathcal{L} ^{(\mathrm{gl})} $ in Eq.~\myref{eq:L_gl} that converts the $N$-node states to a global state $y^{\mathrm{(gl)}} = \mathcal{L} ^{\mathrm{(gl)}} (\bm{x})$, which enables us to compare dynamics of the simplified system to the original one.
			\begin{equation}
				\mathcal{L} ^{(\mathrm{gl})} ( \bm{x})  = \frac{\bm{1}^T \bm{A} \bm{x}  }{ \bm{1}^T \bm{A} \bm{1}} =  \frac{\langle \bm{k^{\mathrm{out}}}\bm{x} \rangle}{ \langle \bm{k} \rangle } \label{eq:L_gl}
			\end{equation}

			To illustrate the efficiency of this dimension-reduction framework, we first apply it to the mutualistic dynamics embedded in different network structures and compare the node states of the $m$-dimensional system with the original network and the one-dimensional system studied in Ref. \cite{gaoUniversalResiliencePatterns2016} as well. 

\noindent
\subsection{The stable state approximation}

We use Eq.~\myref{eq:mutualistic} to model the deterministic dynamics of the mutualistic system, which tracks the abundance of its species \cite{hollandPopulationDynamicsMutualism2002}. The self-dynamics $F(x_i)$ defines the growth of each species as following the logistic law with the Allee effect, while the dynamics defined by $G(x_i, x_j)$ accounts for the mutualistic interaction between pairs of species, $i$ and $j$, with the interaction strength $A_{ij}$ defined in Eq.~\myref{eq:multi-variable}. 
\begin{equation}
\begin{split}
	&F(x_i) = B_i   + x_i  \left(  1- \frac{x_i}{ K_i} \right) \left(\frac{x_i}{ C_i} - 1 \right) \\
	&G(x_i, x_j ) = \frac{x_i x_j }{D_i + E_i x_i + H_j x_j} \label{eq:mutualistic}.
\end{split}
\end{equation}

Note that we use the same parameters as Ref. \cite{gaoUniversalResiliencePatterns2016}. The parameters are node-uniform and set as $B_i = B = 0.1 $, $C_i = C = 1$, $D_i = D = 5$, $E_i = E = 0.9$,  $H_j = H = 0.1$, $K_i = K = 5$. As reported by Ref. \cite{gaoUniversalResiliencePatterns2016}, such dynamics exhibit alternative stable states, the high-stable state, and the low-stable states, which depend on the initial states and coupling strength.

	\begin{figure}
	\centering
		\includegraphics[width=1.0\textwidth]{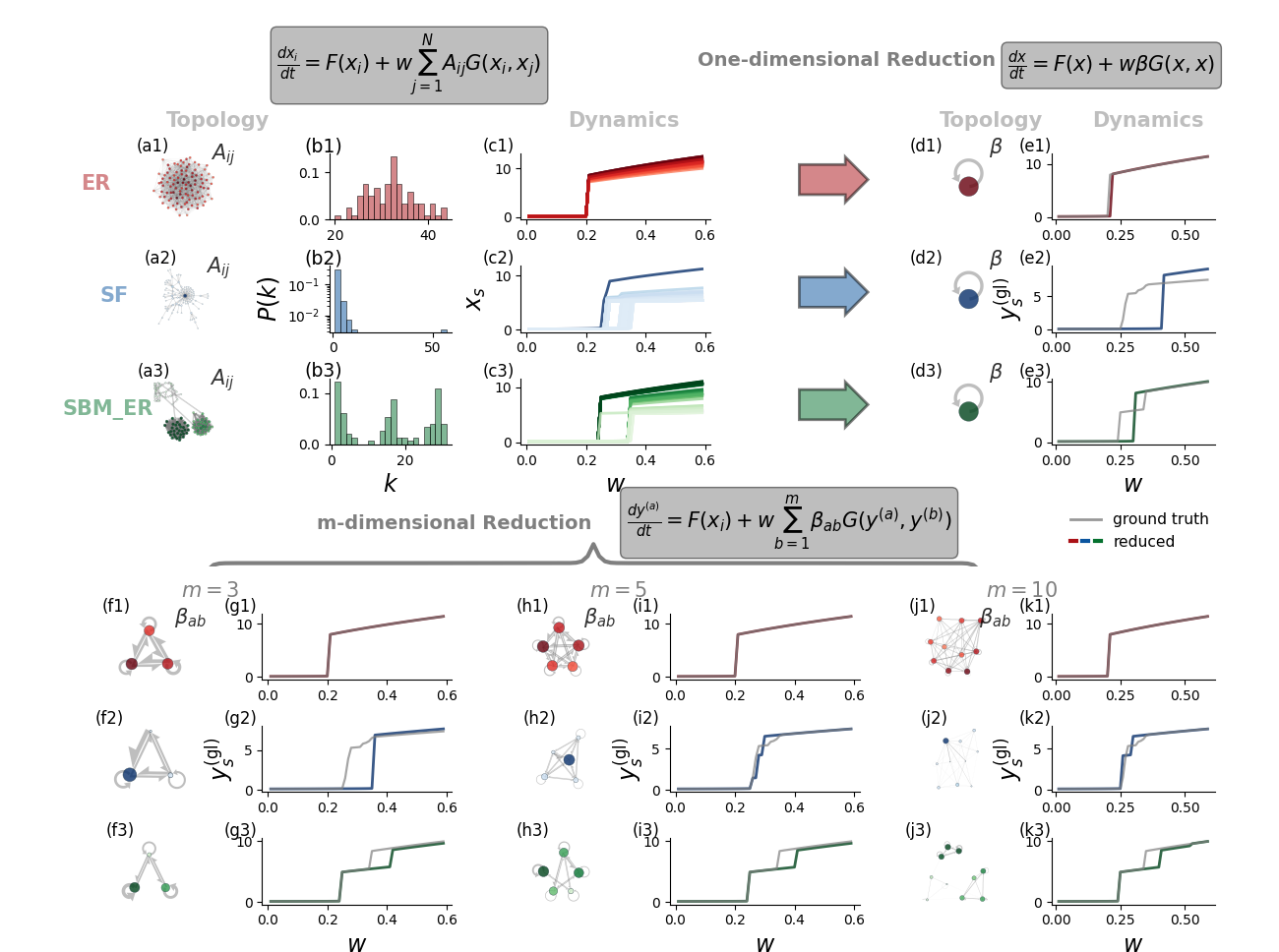}
		\caption{\textbf{The illustration of the network partition and system dynamics defined by the generalized dimension reduction approach.} The system follows mutualistic dynamics. We consider three types of network structures, (1) ER in red ($\langle k \rangle =32$), (2) SF in blue (the power-law exponent $\gamma=2.1$,  the minimal degree $k_{\mathrm{min}}=1$), (3) SBM-ER in green (link probabilities for each community are $p_1=0.9$,  $p_2=0.5$, $p_3=0.05$, while between communities, it is $q=0.001$). Each network consists of $N=100$ nodes. We show the equilibrium states under different dimension-reduction strategies. (a) The topology of the original network, where the node color transparency and size are proportional to the node degree. (b) The degree distribution of the original network. (c) The stable state changes with the edge weight $w$ for all individual nodes. (d) The topology of the one-dimensional system. (e) The stable state changes with the edge weight $w$ for the single node. (f), (h), (j) show the topology of the $m$-dimensional system ($m=3, 5, 10$ respectively), and (g), (i), (k) are the corresponding global stable states versus edge weights $w$. The grey curves are the global state obtained from the numerical solution of the original network. } \label{fig:illutration_mutual}
	\end{figure}

	We study three types of network topology, SF, ER, and SBM-ER (ER network with communities constructed by the stochastic block model \cite{hollandStochasticBlockmodelsFirst1983}) and the phase diagram of the stable states for dimension-reduced systems with different values of $m$ in Fig.~\ref{fig:illutration_mutual}. The equilibrium states of each node in Fig.~\ref{fig:illutration_mutual} c are more diverse for the heterogeneous network and the modular network than the homogeneous ER network, and this partly verifies the correlation between the node states and node degrees. As expected, the one-dimensional reduction system cannot accurately represent the average state of the heterogeneous network, especially when the system is close to the tipping point of phase transition. The tipping point is the critical value of a system parameter, in this case, the edge weight $w$, at which the system undergoes a substantial change in its state (A more rigorous definition of the tipping point is provided in Section II C). As illustrated in Fig.~\ref{fig:illutration_mutual} e, g, i, k, where the system exhibits the transition from low-stable state to high-stable states. In contrast, the one-dimensional reduction system produces reasonable approximations for the homogeneous ER network. Most importantly, it can also capture the critical transition and threshold. In Fig.~\ref{fig:illutration_mutual} f -- k, $m$-dimensional reduction systems are constructed by the degree-based network partition and degree-weighted mean-field approach. The simplified network topology and phase transition of the global state versus the interaction strength are compared with different values of $m$ and complexity. For an SF network with $N=100$ nodes, the state evolution of the five-dimensional reduction system is very close to the ground truth. For the SBM-ER network with three communities, the three-dimensional system captures two substantial changes in the system state out of three, and the global state is accurately predicted when the system is not close to the critical thresholds. One can notice that there is no significant improvement with more clusters divided, which indicates that the community structure, apart from degree attributes, may impact the performance of the dimension-reduced system. 

\begin{figure}
	\centering
		\includegraphics[width=0.95\textwidth]{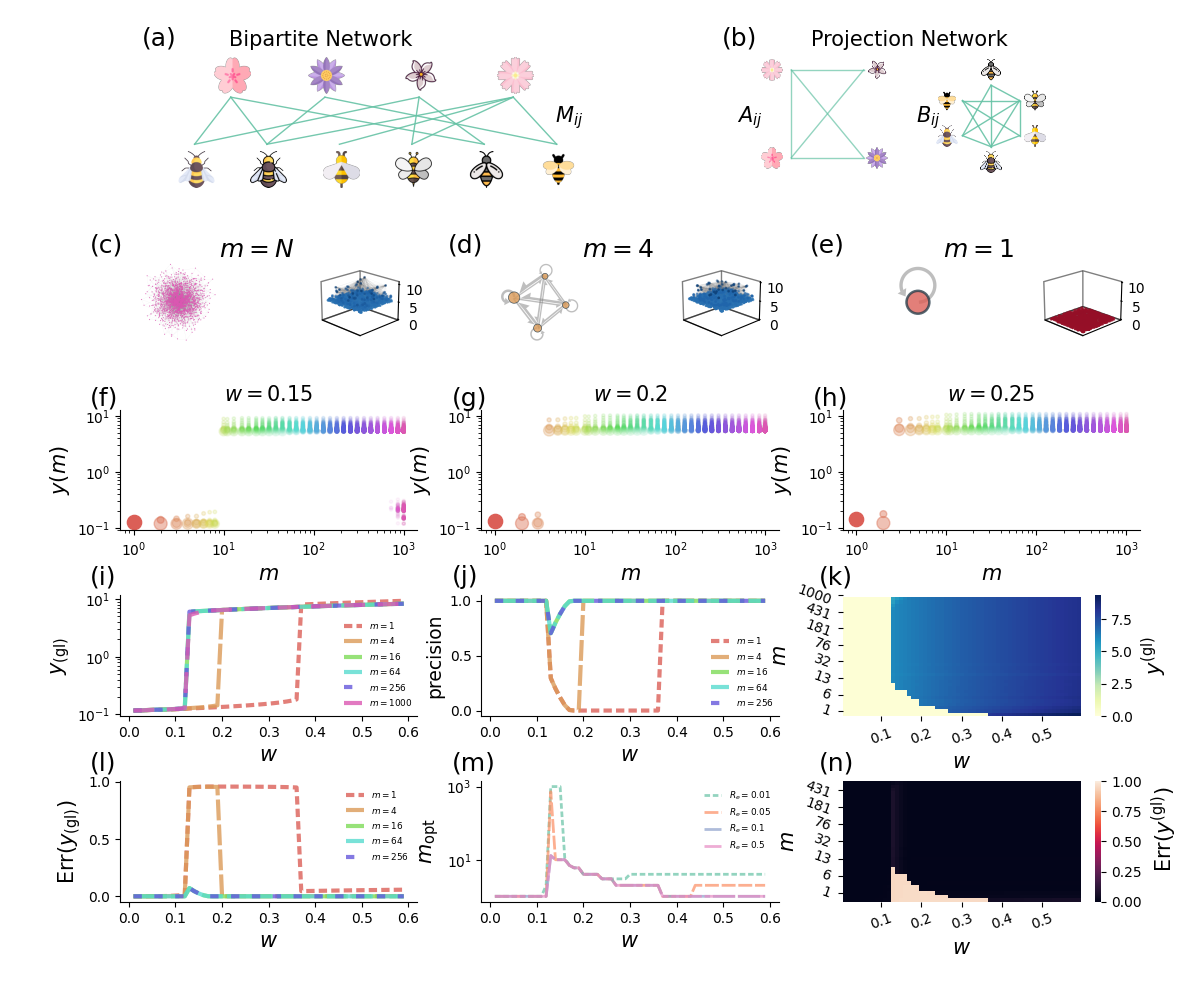}
		\caption{\textbf{A SF network with $N=1000$ nodes following mutualistic dynamics.}  The entire system starts from the low states, $x_i(t=0)=0.1$. (a) The bipartite network $M_{ij}$ describes the connection between pollinators and plants. (b) From $M_{ij}$, we construct two mutualistic projection networks by connecting pairs of plants that feed the same pollinators ($A_{ij}$) and pollinators which share the same plants ($B_{ij}$).  (c) -- (e) The network topology and the stable states for the original network ($m=N$), the low-dimensional reduction system ($m=4$), and the one-dimensional reduction system ($m = 1$) with the edge weight $w=0.2$. For different edge weights (f) $w=0.15$, (g) $w=0.2 $, (h) $w=0.25$, the stable states of each cluster are exhibited against the number of clusters $m$. The marker size is proportional to the number of nodes in that cluster. (i) The global state $y^{(\mathrm{gl})}$ changes with the edge weight $w$ for different values of $m$. (j) The ratio of the qualitatively correct prediction, which counts the fraction of nodes at either high-stable states or low-stable states predicted by both the original network and the dimension-reduced system. The state threshold separating the two stable states is set as $R_y=1$. (k) The heatmap of the global state $y^{(\mathrm{gl})}$ as a function of $w$ and $m$. (l) The relative error of the stable state for different values of $m$ compared to the ground truth. (m) The optimal $m$ changes with $w$ for different values of thresholds $R_e$, and $m_{\mathrm{opt}}$ is defined as the minimal value of $m$ that produces the error smaller than the threshold $R_e$. (n) The heatmap of relative errors of global states compared with the ground truth.} \label{fig:mutual_SF_onenet}
	\end{figure}

	Next, we examine the dimension-reduction framework for a larger SF network in depth and evaluate the approximation accuracy of the state evolution for different coupling strengths and different values of $m$. In Fig.~\ref{fig:mutual_SF_onenet} c -- e, the uniform interaction strength is set as $w=0.2$ for the original network, and all nodes stabilize at the active (high-stable) state. However, the one-dimensional reduction system predicts that the entire system evolves to the undesired (low-stable) state, which is a misrepresentation of the original system, in contrast with the successful prediction of the four-dimensional reduction system. As the interaction strength $w$ increases in Fig.~\ref{fig:mutual_SF_onenet} f -- h, one can use the system with smaller $m$ to qualitatively capture the original system state (whether the system is in the high-stable state or in the low-stable state). This is because the system moves away from the tipping point of phase transition. It can be verified by Fig.~\ref{fig:mutual_SF_onenet} i -- k, which displays the phase transition of the global state and the qualitative precision against the interaction strength. As the coupling strength approaches the tipping point of the original system, larger $m$ is required to qualitatively capture the system state. 
	To quantitatively evaluate the approximation accuracy of this formalism, we compare the global state of the dimension-reduced system with the original network and introduce the relative error of the global state in Eq.~\ref{eq:error_ygl}. In addition, one can set an error threshold $R_e$ to determine the optimal $m$, which is defined as the smallest number of clusters required to produce the relative error less than $R_e$ with the assumption that the prediction accuracy increases with $m$. In Fig.~\ref{fig:mutual_SF_onenet} l -- m, the relative error of the global state increases as the interaction strength approaches the tipping point for a fixed $m$, and the optimal $m_{\mathrm{opt}}$ increases in a similar fashion.
	Fig.~\ref{fig:mutual_SF_onenet} shows the performance of the dimension-reduction theory for one SF network, from which one can observe that the approximation accuracy strongly depends on the network coupling strength (the position in the phase diagram). 

\begin{equation}
	\mathrm{Err(y^{(\mathrm{gl})})} = \frac{|y^{(\mathrm{gl}, N)} - y^{(\mathrm{gl}, m)}|}{|y^{(\mathrm{gl}, N)}| + |y^{(\mathrm{gl}, m})|} \label{eq:error_ygl}
\end{equation}

\begin{figure}
	\centering
		\includegraphics[width=0.9\textwidth]{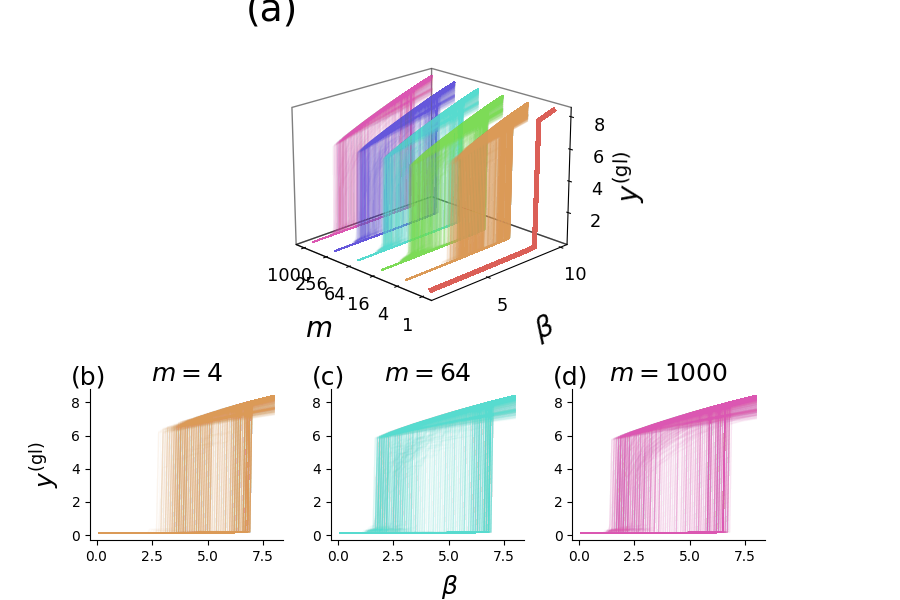}
		\caption{\textbf{The phase diagram of the global state for SF networks following the mutualistic dynamics.} The stable states are obtained from $L=900$ networks (the exponent $\gamma$ ranges from $2.1$ to $5$, and the minimal degree $k_{min}$ is $3$, $4$, or $5$), and all of them start from the low states, $x_i(t=0)=0.1$. (a) shows the phase diagram for different values of $m$. The parameter $\beta$ is calculated by the one-dimensional reduction theory \cite{gaoUniversalResiliencePatterns2016}. In each subplot (b) -- (d), the phase diagram of the global state against the parameter $\beta$ for the same system dimensionality (the same $m$) and each curve represents one individual network. } \label{fig:mutual_SF_ygl_beta}
	\end{figure}

For SF networks with different heterogeneities, one can observe the phase diagram of the original network and dimension-reduced systems are different from the one-dimensional system. When we include more components (larger $m$), the equilibrium state approximation improves in Fig.~\ref{fig:mutual_SF_ygl_beta}. 
For the systems far away from the tipping point of phase transition, one-dimensional systems ($m=1$) are already sufficient to represent the system state.  

To systematically validate the proposed framework, we apply it to two more dynamical models, neuronal dynamics \cite{wilsonExcitatoryInhibitoryInteractions1972, wilsonMathematicalTheoryFunctional1973} and gene regulatory dynamics \cite{alonIntroductionSystemsBiology2006, karlebachModellingAnalysisGene2008}, which also exhibit phase transitions between alternative stable states. 
The Wilson-Cowan neuronal dynamics in Eq.~\myref{eq:neuronal} describes the firing-rate activity of a population of neurons. 
Although the Wilson-Cowan equations are already an approximation by mean-field theory \cite{buiceMeanFieldTheory2013, kuehnMomentClosureBrief2016, bickUnderstandingDynamicsBiological2020}, they can still exhibit very complicated behavior for the large system. Our approach provides an alternative and complementary way to simplify the neural dynamics.
\begin{equation}
	\dv[]{x_i}{t} = -x_i + \sum _{j=1}^{N}A_{ij} \frac{1}{1+ e^{\theta - \delta x_j}}, \label{eq:neuronal}
\end{equation}
where the parameters $\theta=5$ and $\delta=1$ control the firing-rate threshold and the steepness of the activation function, respectively. 

We also apply our approach to biological networks. The gene regulatory dynamics are governed by the Michaelis-Menten equation \myref{eq:genereg}:
\begin{equation}
	\dv[]{x_i}{t} = -B x_i^f + \sum _{j=1}^{N} A_{ij}  \frac{x_j ^ h }{x_j ^ h + 1},	\label{eq:genereg}
\end{equation}
where the parameters are also node-uniform and set as $B = 1$, $f=1$, $h=2$. The first term on the right-hand side describes the degradation of each gene, and the second term captures genetic activation, indicating the level of cooperation between genes.

\begin{figure}
	\centering
		\includegraphics[width=0.9\textwidth]{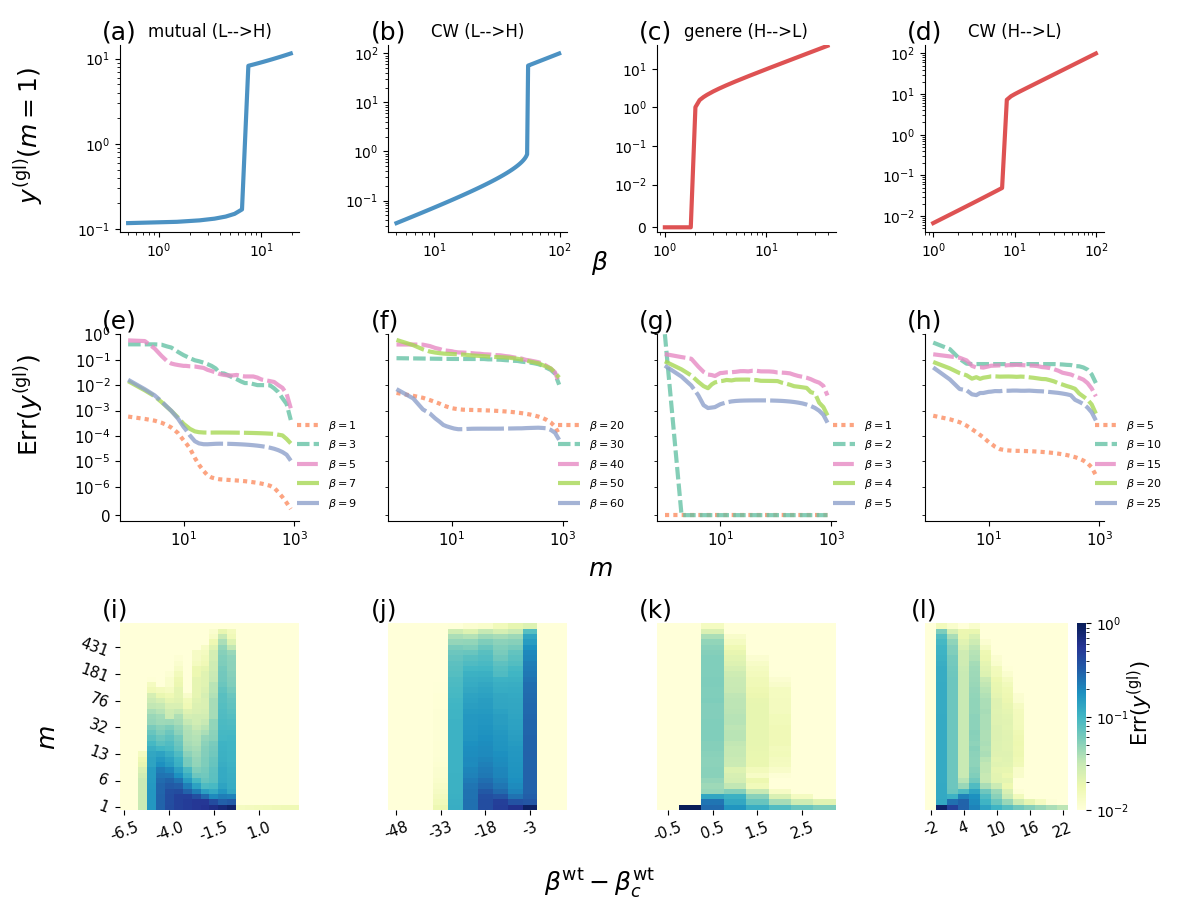}
		\caption{\textbf{The relative error of the global state for different dynamics.}  Evolution data is collected from $L=900$ networks (the exponent $\gamma$ ranges from $2.1$ to $5$, and the minimal degree $k_{min}$ is $3$, $4$, or $5$). (a) -- (d) display the phase diagram of the stable states against the effective interaction strength $\beta$, which is calculated by a single-dimension reduction system. For (a) mutualistic dynamics starting from the low state $x_i(t=0) = 0.1$, the tipping point of phase transition $\beta_c^{(m=1)}=7$, (b) Wilson-Cowan (CW) neuronal dynamics starting from the low state $x_i(t=0) = 0.0$, $\beta_c^{(m=1)}=56$, (c) gene regulatory dynamics starting from the high state $x_i(t=0) = 100$, $\beta_c^{(m=1)}=2$, and (d) CW neuronal dynamics starting from the high state $x_i(t=0) = 100$, $\beta_c^{(m=1)}=8$. (e) -(h) The error of the global state is calculated for $m$-dimensional systems in the comparison of the original networks. (i) -- (l) are heatmaps of the error as a function of dimensionality $m$ and the distance to the tipping point $\beta_c^{(m)}$.  } \label{fig:SF_yerror_beta_m}
	\end{figure}

In Fig.~\ref{fig:SF_yerror_beta_m}, one can observe that the prediction accuracy depends on the coupling strengths. For different dynamics, the relative error of $y^{\mathrm{(gl)}}$ behaves in a similar fashion. As the system interaction strength approaches the tipping point of the one-dimensional system, the error increases in most cases.  Therefore, more clusters are needed to correctly understand the system states.

\noindent
\subsection{Tipping point approximation}

	For heterogeneous networks, the $m$-dimensional reduction strategy achieves a much better approximation accuracy than the one-dimensional reduction model. In many cases, a small number of clusters can capture the system evolution. The only exception is that as the system approaches the tipping point of phase transition, the number of clusters should be increased to help qualitatively predict the system state and identify critical thresholds. Fig.\ref{fig:SF_yerror_beta_m} also shows that the prediction accuracy depends on the distance to the tipping points. 
	Hence, locating the tipping point of phase transition is important for distinguishing the active state from the dead state.
	 To determine the critical point of either the dimension-reduced system or the original networked system, appropriate criteria need to be chosen. Here, we introduce two types of thresholds, the survival ratio threshold $R_s$ (nodes with a state above a certain value are considered to be in a surviving state), and the global state threshold $R_y$. The tipping point $w_c$ is defined as the smallest interaction strength at which the ratio of survival nodes or the global state exceeds the predefined threshold ($R_s$ or $R_y$).
	 In addition, one can quantify the approximation accuracy by calculating the relative difference of the tipping points between the dimension-reduced system and the original network using Eq.~\myref{eq:wc_N_m}. To demonstrate the improvement of including more than one component, the normalized distance of the critical point in $h^{\mathrm{wt}} - \langle k \rangle ^{\mathrm{wt}} $ space is introduced in Eq.~\myref{eq:l_m_1}, which is used to represent the difference of thresholds between the one-dimensional reduction system and the $m$-dimensional system. Here, $h^{\mathrm{wt}}$ is the weighted degree heterogeneity defined as $h^{\mathrm{wt}} = wh   = w \frac{\langle k^2 \rangle  - \langle k \rangle ^2}{ \langle k \rangle } $,  $\langle k \rangle ^{\mathrm{wt}} = w\langle k \rangle $ is the weighted average degree, and $\beta^{\mathrm{wt}} = w \frac{\langle k^2 \rangle }{ \langle k \rangle } = h^{\mathrm{wt}} - \langle k \rangle ^{\mathrm{wt}}$ is the weighted effective interaction strength of the one-dimensional reduction system. 
	
\begin{figure}
	\centering
		\includegraphics[width=1.0\textwidth]{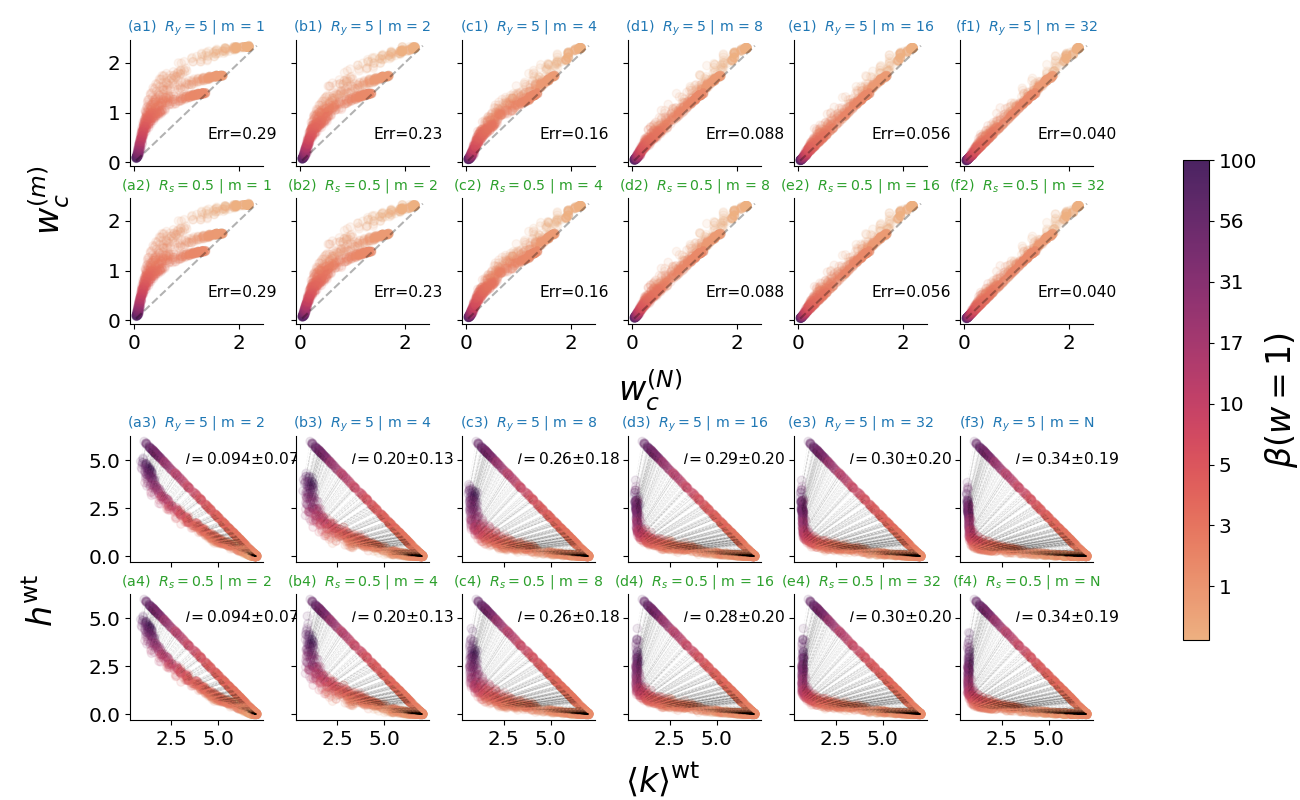}
		\caption{\textbf{The tipping point approximation of the dimension-reduction framework.} The results are obtained from $L = 900$ SF networks. (a1) -- (f2) compare the tipping points between dimension-reduced systems and the original networks. (a3) -- (f4) are the comparisons of the tipping points between multi-dimensional systems and one-dimensional systems in $h^{\mathrm{wt}} - \langle k \rangle ^{\mathrm{wt}}$ space. Two error thresholds, $R_y$ and $R_s$, measure the difference of the global state and the survival ratio, respectively. } \label{fig:mutual_SF_wc_beta_hk}
	\end{figure}

\begin{figure}
	\centering
		\includegraphics[width=1.0\textwidth]{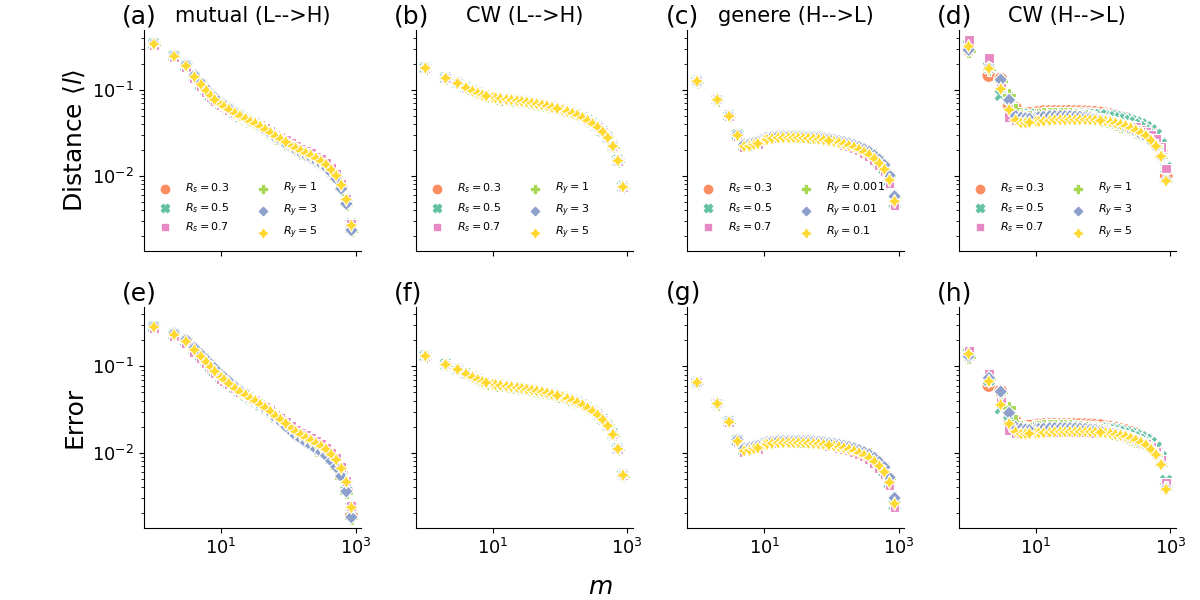}
		\caption{\textbf{ The difference of the critical point changes with the number of clusters for SF networks with different dynamics.}  For each dynamics, the results are obtained from $L = 900$ SF networks. Different symbols represent different threshold types and values. }  \label{fig:mutual_SF_dis_err}
	\end{figure}

	In the one-dimensional reduction framework, all networks with different degree distributions share the universal tipping point $\beta^{\mathrm{wt}}_c$ if they follow the same dynamical mechanism \cite{gaoUniversalResiliencePatterns2016}. As we have already noticed in Fig.~\ref{fig:illutration_mutual} and \ref{fig:mutual_SF_onenet}, the tipping points of heterogeneous networks depend on network topology and they cannot be accurately captured by the one-dimensional systems. In Fig.~\ref{fig:mutual_SF_wc_beta_hk}, we quantitatively compare the tipping points between the $m$-dimensional reduction system and the original network and demonstrate that the difference typically decreases as $m$ increases. We also show that the tipping points predicted by the one-dimensional reduction system significantly deviate from the ground truth, and the difference depends on the network topology. From Fig.~\ref{fig:mutual_SF_dis_err}, one can observe that the average accuracy of SF network ensembles increases as the number of clusters increases, and such results hold for different types of dynamics. 

	\begin{equation}
		\mathrm{Err} = \frac{|w_c^{(m)} - w_c^{(N)}|}{w_c^{(m)} + w_c^{(N)}} \label{eq:wc_N_m}
	\end{equation}

	\begin{equation}
		l = |w_c^{(m)} - w_c^{(N)}| \frac{ \sqrt{\langle k \rangle ^2 + h ^2 } }{ \beta_c^{wt}} \label{eq:l_m_1}
	\end{equation}

\section{The application to time-delay systems}

	In this section, we use this framework to study the dynamical networks in the presence of time delays and further analyze the critical time delay in terms of stability. 
	
Delayed interactions are quite common in many dynamical systems and have been extensively studied in terms of stability and synchronization \cite{olfati-saberConsensusCooperationNetworked2007, erneuxAppliedDelayDifferential2009, olfati-saberConsensusCooperationNetworked2007, huntNetworkSynchronizationNoisy2010, huntNetworkCoordinationSynchronization2012, huntImpactCompetingTime2011, huntExtremeFluctuationsStochastic2015, ottoNonlinearDynamicsDelay2019, piganiDelayEffectsStability2022}. The key early results were published by Hutchinson \cite{hutchinsonCircularCausalSystems1948} and May \cite{mayTimeDelayStabilityPopulation1973}, who discussed the time delay and stability in the single-variable ecological systems. \cite{hutchinsonCircularCausalSystems1948, mayTimeDelayStabilityPopulation1973, ruanAbsoluteStabilityConditional2001, ruanDELAYDIFFERENTIALEQUATIONS2006}. As the time delay occurring on one individual has profound effects on its neighboring components in the interconnected systems, one cannot capture the collective phenomena or global dynamics by only investigating one or two components. Therefore, one might need to investigate the complex interaction topology to understand the system evolution and stability. On the other hand, serious computational issues arise for simulating large networked systems. 
	In this case, one can resort to the dimension-reduction framework to reduce computational complexity while preserving key information about the interactions between components.

	The network dynamics in the presence of delays can be mathematically characterized by the delay differential equations (DDE). In this study, we focus on the dynamics with constant time delay and the delay effects are incorporated into the self-interaction term. Correspondingly, the time-delayed dynamics of the system composed of $N$ interacting nodes is described by Eq.~\myref{eq:dde}. As before, the term ``dimension" refers to ``the number of nodes" in the original system or ``the number of clusters" in the simplified system.
	\begin{equation}
		\dv[]{x_i}{ t} = f_i(x_i, x_i(t-\tau), x_j) = F(x_i, x_i(t-\tau)) +w \sum _{j=1}^{N} A_{ij}G(x_i, x_j)   \label{eq:dde}
	\end{equation}

	We apply the degree-based dimension-reduction approach to the delayed system and obtain the $m$-dimensional system  defined by Eq.~\myref{eq:dde_dr}, where $y_a$ is the average node state of the cluster $g_{a}$, and $\beta_{ab}$ is the interaction strength from the cluster  $g_b$ to the cluster $g_a$.
	\begin{equation}
		\dv[]{y_a}{ t} = f_a(y_a, y_a(t-\tau), y_b) = F(y_a, y_a(t-\tau)) +w \sum _{b=1}^{m}\beta_{ab}G(y_a, y_b)   \label{eq:dde_dr}
	\end{equation}

\begin{figure}
	\centering
		\includegraphics[width=0.95\textwidth]{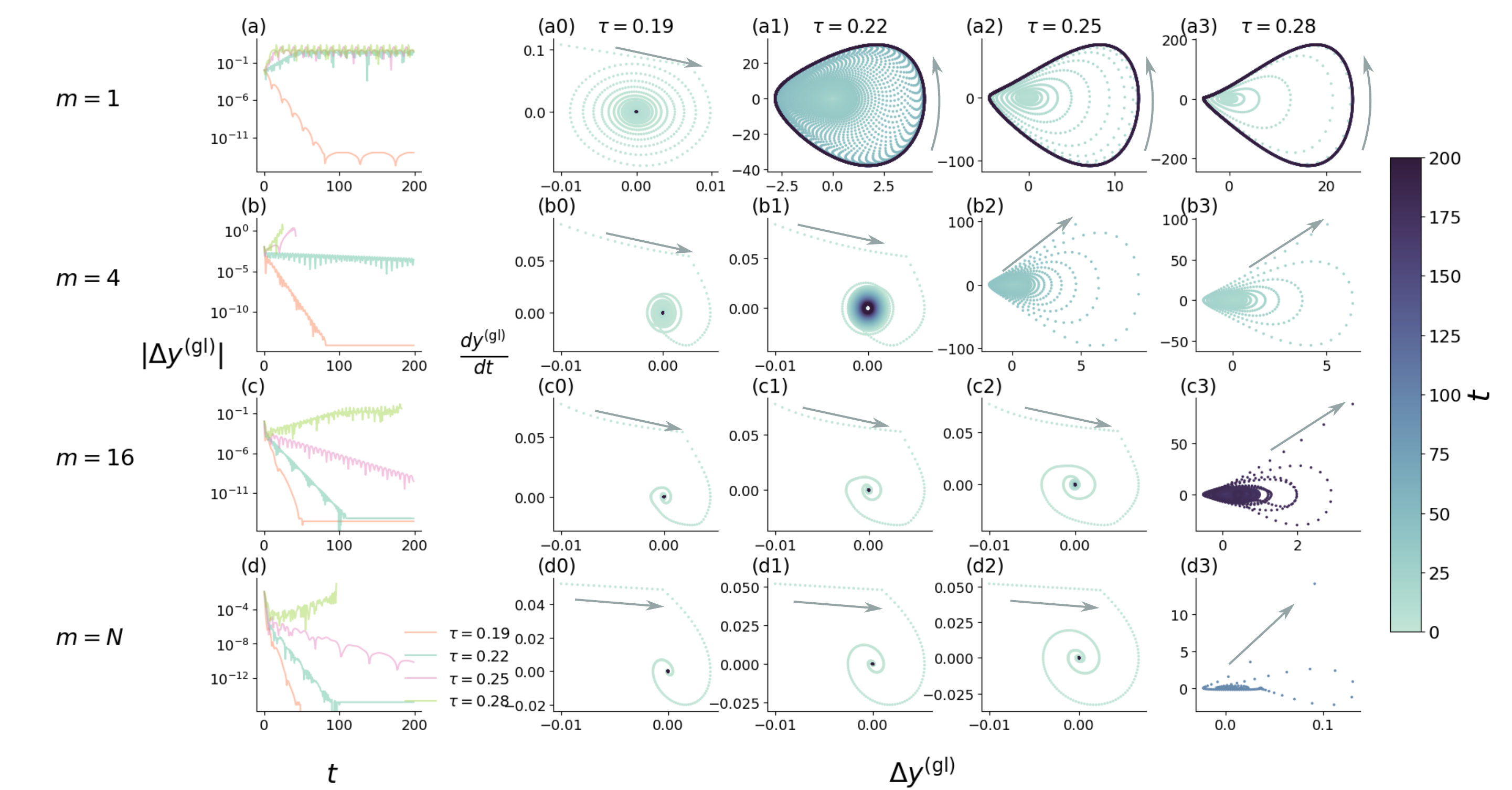}
		\caption{\textbf{A SF network of $N=1000$ nodes and its corresponding dimension-reduced systems under different time delays.} This SF network is constructed by the configurational model with the exponent parameter $\gamma=2.5$, and the minimal degree $k_{\mathrm{min}}=3$. The edge weight $w=0.1$. (a) -- (d) The evolution of the global state difference $|\Delta y^{\mathrm{(gl)}}|$ for different values of time delay $\tau = 0.19, 0.22, 0.25, 0.28$ and the system dimension $m = 1, 4, 16, N$. $\Delta y^{\mathrm{(gl)}}$ is the difference between the system global state at time $t$ and the stable state without time delay. Other figures show the phase space $\dv[]{y^{\mathrm(gl)}}{t}$ versus $\Delta y^{\mathrm(gl)}$respectively.  } \label{fig:mutual_one_SF_tau_phase_space_w=0.1}
	\end{figure}

\begin{figure}
	\centering
		\includegraphics[width=0.95\textwidth]{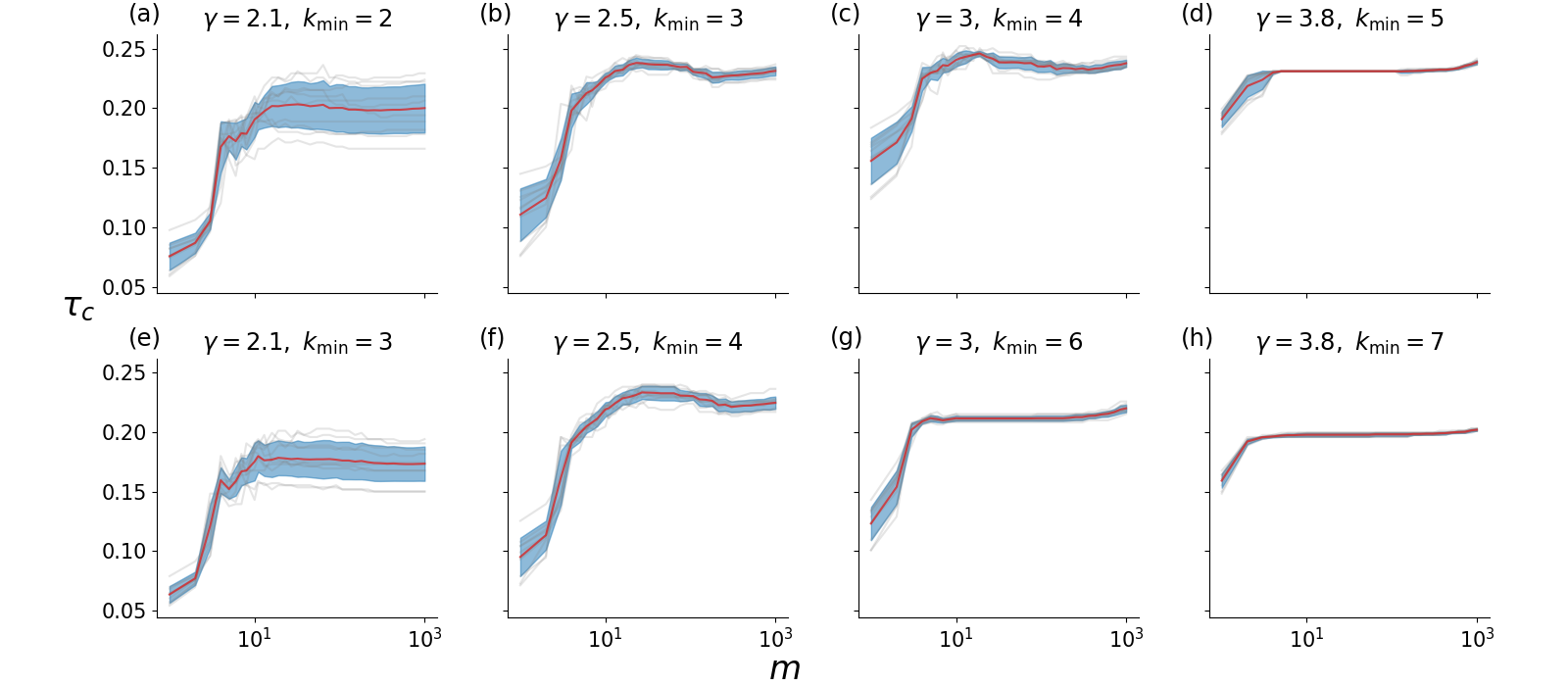}
		\caption{\textbf{The critical time delay $\tau_c$ determined by dynamical evolution for SF networks using the dimension-reduction approach.} All networks are created by configuration model and they consist of $N=1000$ nodes. For each parameter choice, there are $10$ realizations. The edge weights are the same, $w=0.6$. (a) -- (d) The average degree $\langle k \rangle  \sim 6$, and (e) -- (h) $\langle k \rangle \sim 10$. } \label{fig:mutual_SF_tau_c_evolution_w=0.6}
	\end{figure}

\begin{figure}
	\centering
		\includegraphics[width=0.95\textwidth]{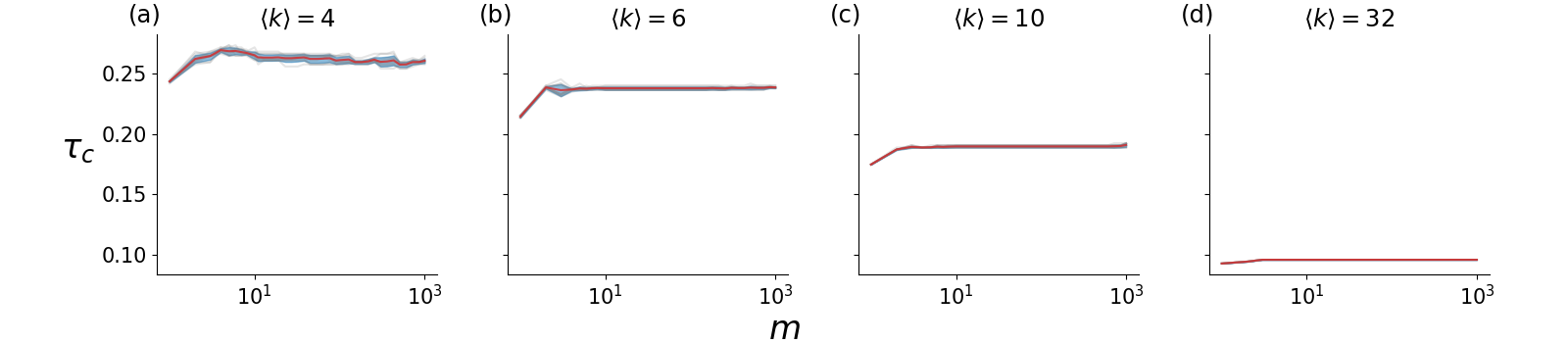}
		\caption{\textbf{The critical time delay $\tau_c$ determined by dynamical evolution for ER networks using the dimension-reduction approach.} All networks are created by configuration model and they consist of $N=1000$ nodes. The edge weights are the same, $w=0.6$. For each figure, there are $10$ realizations. } \label{fig:mutual_ER_tau_c_evolution_w=0.6}
	\end{figure}

	Fig.~\ref{fig:mutual_one_SF_tau_phase_space_w=0.1} shows the impact of time delays on the dynamical stability for a SF network and its corresponding dimension-reduced versions. As the time delay increases, the system takes longer time to converge to the stable state. If the time delay exceeds a certain threshold, the system will either go to the limit cycle for the low-dimension systems (Fig.~\ref{fig:mutual_one_SF_tau_phase_space_w=0.1} a1--a3) or diverge (Fig.~\ref{fig:mutual_one_SF_tau_phase_space_w=0.1} b2 b3, c3, d3), so that one can determine the critical time delay in terms of stability based on the system evolution. One can notice that the low-dimensional system ($m=1$ or $m=4$) does not only yield quantitative deviations from the ground truth, but can also qualitatively miss the actual behavior [limit cycle vs. stationary value], as shown in Fig.~\ref{fig:mutual_one_SF_tau_phase_space_w=0.1}. One needs to employ a higher-$m$ ($m=16$ in this example) reduction scheme to at least qualitatively capture the nature of the steady state, and progressively higher ones for further quantitative improvements.
To determine this critical threshold, one can numerically simulate the system evolution under different values of time delays. By using the binary search algorithm, one can find the largest time delay with which the system still converges to a steady stable. For large networked systems, this method requires substantial computational resources. Hence, one can instead use the dimension-reduction framework to reduce the system complexity. One can observe that the global state evolution becomes more accurate compared to the ground truth (the evolution of the original system) as the number of clusters increases (Fig.~\ref{fig:mutual_one_SF_tau_phase_space_w=0.1} c and d). 
	Furthermore, the critical time delay can be predicted by the dimension-reduced systems for different network topologies as shown in Fig.~\ref{fig:mutual_SF_tau_c_evolution_w=0.6} and \ref{fig:mutual_ER_tau_c_evolution_w=0.6}, and the error decreases dramatically as more clusters are included. The more heterogeneous the network is (smaller $\gamma$ for SF networks), the larger number of clusters may be needed. This approach proves to be very efficient since even for very heterogeneous networks, the system with $m=10$ clusters can produce reasonable approximations.

	Alternatively, one can derive the characteristic equation \myref{eq:det_M} \cite{ruanAbsoluteStabilityConditional2001, sipahiStabilityStabilizationSystems2011} associated with the system \myref{eq:dde} by exerting a small perturbation $\bm{R} = \bm{c} e^{\alpha t}$ ($\alpha=\mu + i \nu$) around the equilibrium and performing the first order approximation to obtain the linear stability 
	\begin{equation}
		\det(\bm{M}(\alpha; \tau)) = \det(\alpha \bm{I} - \bm{J}_{\bm{f}} (\bm{x}) \lvert _{\bm{x}_s}- \bm{J}_{\bm{f}} (\bm{x}_{\tau}) \lvert _{\bm{x}_s} e^{-\alpha \tau}) = 0 \label{eq:det_M}
	\end{equation}

	 In Eq.~\myref{eq:det_M}, $\bm{I}$ is the  identity matrix, $\bm{J}_{\bm{f}} (\bm{x})$ and $\bm{J}_{\bm{f}} (\bm{x}_{\tau})$ are the Jacobian matrices, $\bm{f} = (f_1, f_2, ..., f_N) ^T$ is a vector of individual dynamical functions, $\bm{x}= (x_1, x_2, ..., x_N)^T$ is a vector of node states at time $t$, and $\bm{x}_{\tau}=(x_1(t-\tau), x_2(t-\tau), ...,  x_N(t-\tau))$ is a vector of node states with delay $\tau$ (i.e., at time $t-\tau$). The (i, j) element of matrix $\bm{J}_{\bm{f}}(\bm{x})$, for example, is calculated by $\pdv[]{f_i}{x_j}$.
The system is asymptotically stable if and only if all roots $\alpha$ of Eq.~\myref{eq:det_M} have negative real parts \cite{ruanAbsoluteStabilityConditional2001}. When the time delay reaches the threshold $\tau_c$, Eq.~\myref{eq:det_M} has a pair of purely imaginary roots ($ \pm i \nu$), which is the critical condition. Hence, for any delay $\tau < \tau_c$, the real parts of all roots $\alpha$ remain negative, leading to asymptotically stable convergence, and when $\tau > \tau_c$, there is at least one root with positive real part, which destabilizes the system, according to Rouche's theorem and continuity \cite{dieudonneFoundationsModernAnalysis2008}. 

\begin{figure}
	\centering
		\includegraphics[width=0.95\textwidth]{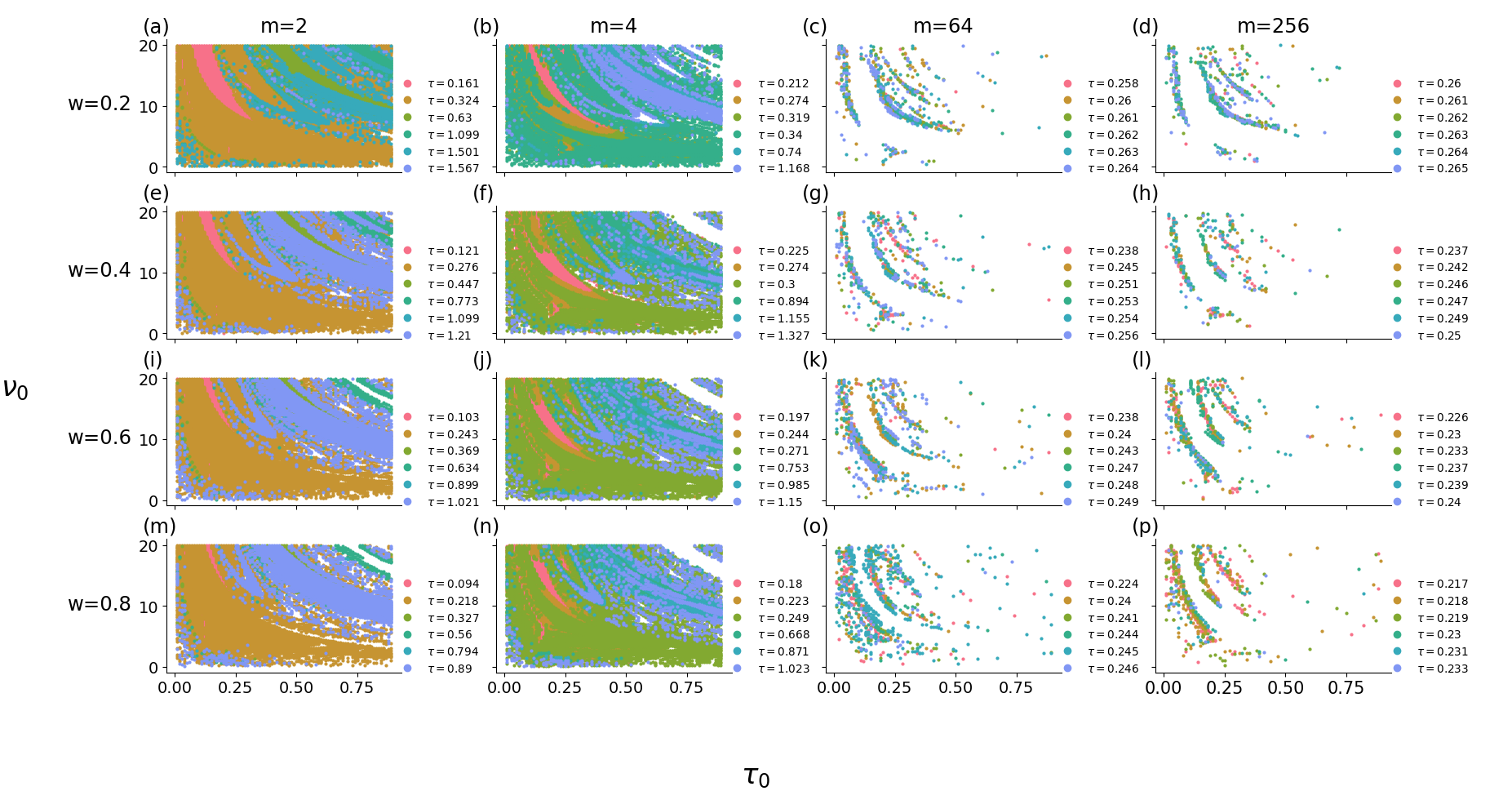}
		\caption{\textbf{Solutions to Eq.~\myref{eq:det_M} for different initial parameters $(\tau_0, \nu_0)$.} The system consists of $N=1000$ nodes. Results are for different edge weights $w$ and system dimensionality $m$.} \label{fig:mutual_SF_one_tau_initial}
	\end{figure}

\begin{figure}
	\centering
		\includegraphics[width=0.95\textwidth]{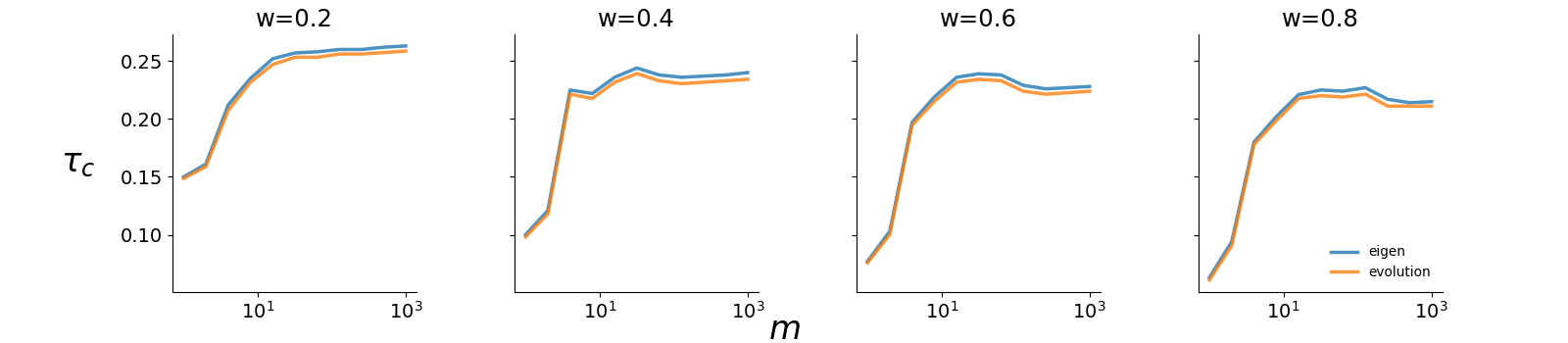}
		\caption{\textbf{The comparison of the critical time delay obtained from the system evolution and the characteristic equation.} There are $N=1000$ nodes in the system. } \label{fig:mutual_SF_tau_eigen_evo_compare}
	\end{figure}

To determine the critical time delay, one should first set $\alpha=i\nu$, and then solve the characteristic equation \myref{eq:det_M}. Because it is essentially a polynomial equation, there are multiple solutions. For the critical condition, only the smallest positive solution is of interest. It turns out to be another challenging numerical problem as the solutions strongly depend on the initial condition shown in Fig.~\ref{fig:mutual_SF_one_tau_initial}. The heuristic approach is to increase the search space of $(\tau_0, \nu_0)$ in order to find out more solutions and choose the smallest one as the critical point. Although the solution obtained by this method is not guaranteed to be the ground truth, it demonstrates the accuracy in comparison to the result from the numerical evolution in Fig.~\ref{fig:mutual_SF_tau_eigen_evo_compare}.
In addition, the critical time delays obtained by the dimension-reduced systems are in good agreement with the original system as long as a sufficient number of clusters ($m=10$) are included.

\section{Discussions}
In this study, we present a mean-field framework to reduce the dimensionality of large complex networks. In this framework, the network is first classified into $m$ clusters based on node degrees such that the nodes in the same cluster share similar degrees with the assumption that nodes of similar degrees have similar dynamical evolutions. For each cluster, one observable is then introduced to represent the degree-weighted average state. Therefore, an $N$-dimensional network is described by an $m$-dimensional system and the dimension-reduced version preserves the original dynamical rules. Compared with the previous one-dimensional mean-field approaches \cite{gaoUniversalResiliencePatterns2016, laurenceSpectralDimensionReduction2019}, this framework better approximates the state evolutions and tipping points, especially for heterogeneous networks and networks with communities. We find that the approximation accuracy depends on the network topology, the distance to the underlying tipping point of phase transition, and the number of clusters $m$. Generally, the approximation error increases as the system approaches the tipping point, which indicates that more clusters are needed to have a reasonable understanding of system evolutions. Under different interaction strengths, the number of clusters required to obtain the predefined accuracy (i.e., the value of $m_{\mathrm{opt}}$) differs. It can be determined by comparing the node evolutions of the $m$-dimensional reduction system with the original network. It turns out that $m_{\mathrm{opt}}$ is almost always much smaller than the number of nodes $N$ in the original network. In other words, one can have a reasonable approximation of system evolutions by constructing a much simpler system with $m_{\mathrm{opt}}$ representatives instead of investigating the original network. Another advantage of this dimension-reduction framework is that the interpretation is very clear, as each variable in the dimension-reduced system represents the average of node states in that cluster. Further, one can study how the interactions between clusters influence the cluster states and the system evolution, especially around the tipping points of phase transition. This may guide us to recover the dysfunctional systems by controlling certain clusters \cite{sanhedrai2022reviving,ma2021universality}. 

This framework can not only be used to approximate the system state, but it also has some other potential applications. Here we show that one can apply this theory to time-delay dynamical systems and evaluate the critical threshold in terms of the system stability. In the presence of time delays, the system becomes unstable if the delay is greater than the critical point. Using this method, one can avoid numerically simulating the dynamics of all components or computing the eigenvalues of characteristic matrices for large networks. We also show that our approach can be universally applied to different types of networks and dynamics or with co-adaptation \cite{zhang2020co}. 

Despite the promising results, there are still open questions that need to be addressed in future research. For instance, the impact of network topology and dynamics on the optimal number of clusters remains unclear. Currently, the choice of the number of clusters is determined empirically for each system. This issue also arises when the dimension-reduction framework is applied to time-delay systems. A theoretical investigation of this issue would provide guidelines for the relationship between the error produced by the mean-field approximation and the system dimension. Our approach will be significantly empowered if the method is developed to compute how many clusters the system should be partitioned into based on the network structure and its dynamics from a more theoretical perspective. 

\section*{Data availability statement}
The authors declare that data supporting the findings of this study are available within the paper.  

\section*{Code availability statement}
All codes for the reproduction of the reported results in this study are available upon request.

\bibliographystyle{apsrev4-2_CN}

\bibliography{manuscript_v1}

\end{document}